# Oxygen Chemical Diffusion Coefficient in Manganite Thin Films by Isothermal Electric Resistivity Measurements

*Lorenzo Malavasi*[*a], *and Giorgio Flor*

[a]Dipartimento di Chimica Fisica "M. Rolla", INSTM, IENI/CNR Unità di Pavia of Università di Pavia, V.le Taramelli 16, I-27100, Pavia, Italy.
*E-mail: malavasi@chifis.unipv.it

**Abstract**

In this paper we report the determination of oxygen diffusion coefficient for optimally doped manganite thin films of $La_{0.85}Na_{0.15}MnO_3$ (LNMO). The study was performed by means of isothermal conductivity measurements on two different epitaxially oriented thin films grown on $SrTiO_3$ (100) and $SrTiO_3$ (110). The values found for the oxygen diffusion coefficients are around $10^{-14}$ cm$^2$s$^{-1}$. The diffusion seems to be faster for the *c*-oriented film with respect to the other one (growth on $SrTiO_3$ (110)). The activation energy for the diffusion process is 0.35 eV for the LNMO film on $SrTiO_3$ (100) and 0.66 for the LNMO film on $SrTiO_3$ (110). A possible reason to account for the different behaviour between the two films is proposed.

## Introduction

Mixed valence manganites are highly attractive materials due to the Colossal Magnetoresistive effect (CMR) first observed in Sr-doped $LaMnO_{3+\delta}$[1-4] thin films. The research efforts during these years have pointed out the role played in the manganites by the different oxidation states, (III) and (IV), of the manganese ions: in $LaMnO_{3+\delta}$ and $La_{1-x}Ca_xMnO_3$ systems, for example, a Mn(IV) content around ~30% was found to be the optimal value in order to achieve the maximum CMR effect and also the highest Curie temperature ($T_c$) for the paramagnetic to ferromagnetic transition (P-F)[5-7]. In the magnetic ordered state the electron hopping process is favored by the spin arrangement while for temperatures above $T_c$ the conduction mechanism is semiconducting-like. This behavior was closely related to the double-exchange (DE) mechanism, first proposed by Zener[8], which involves the transfer of $e_g$ electrons between neighboring Mn ions when these are ferromagnetically coupled. Of course, one of the most exciting feature of manganites thin films is the close correlations between the electronic transport and the strain effect induced by the substrate used for thin film growth[9-14].

Another interesting feature of mixed-valence manganites is the role played by oxygen non-stoichiometry on their physical properties. For what concerns the bulk materials there are some thorough works which enabled to figure out a clear picture of the crucial role of oxygen content on the structural, electronic and magnetic properties of these oxides[7,15-17]. Surprisingly, very few reports attempted to study the relationship between oxygen stoichiometry and physical properties of thin films manganites[18-20] even tough annealing treatments are routinely performed on as-deposited thin films. Of course, the basic requirement of a thermal treatment is the certainty of equilibrium attainment. This can be only known a priori if oxygen diffusion coefficient ($D$) is known. Even though the past experience on high temperature superconductors (HTSC) showed that the $D$ values are relatively connected to the nature of the materials and the synthesis conditions, *i.e.* to the specific peculiarities of studied samples, a general knowledge of $D$ magnitude is of crucial importance. Moreover, thin film of manganites are often grown onto substrates with different orientations and the possibility of anisotropies in the oxygen diffusion within the perovskite structure should be taken into account as again observed in the past for HTSC.

In this paper we report the first determination of oxygen diffusion coefficients for $La_{0.85}Na_{0.15}MnO_3$ thin films deposited by means of RF-magnetron sputtering onto $SrTiO_3$ (100) and (110). The study of the oxygen diffusion was carried out by means of isothermal electric resistivity measurements.

**Experimental Section**

La$_{0.85}$Na$_{0.15}$MnO$_{3+\delta}$ thin films were grown starting from powdered target material synthesised by solid state reaction; details about the target preparation and characterisation are reported in Ref 15. The starting perovskite is rhombohedric with lattice constants $a = b = 5.494$ Å and $c = 13.302$ Å. Thin films were deposited on SrTiO$_3$ (STO) (100) and (110) single crystals (Mateck®). The depositions were performed by an off-axis RF-magnetron sputtering system (RIAL Vacuum®); the gas in the chamber was an argon/oxygen mixture in the ratio 12:1. This gas composition was chose since in our previous investigation we could observe that the formation of cation stoichiometric films, with respect to the target material, can be well accomplished in an oxygen-poor gas environment[18]. The total pressure in the sputtering chamber was 4•10$^{-6}$ bar, and the RF power was kept at 145 W. The substrate temperature, measured by a K-type thermocouple located under the substrate, was set at 700°C. Film thickness was monitored by means of an internal quartz microbalance and precisely defined by X-ray reflectivity measurements (XRR).

X-ray reflectivity (XRR), as well as $\theta$-$2\theta$ X-ray diffraction (XRD) and $\omega$ scans data have been collected by using a Bruker "D8 Advance" diffractometer equipped with a Göbel mirror. The Cu K$_\alpha$ line of a conventional X-ray source powered at 40 kV and 40 mA was used. All the reflectivity spectra have been analysed with the REFTOOL software package. X-ray diffraction measurements and XRR results indicate the growth of 50 nm $c$-oriented thin films on the STO (100) single crystal ($a$ and $b$ short lattice parameters in the single crystal plane) while on the STO (110) substrate the film grew with both the $c$-axis and $a$ (or $b$) short axis of the manganite structure in the plane of the film[21,22].

Oxygen diffusion measurements were performed with the DC-four probe method by recording the resistivity ($\rho$) variation as a function of time ($t$), at constant temperature ($T$), when the gas composition was changed between pure oxygen ($P(O_2) = 1$ atm) and $P(O_2) = 10^{-5}$ atm. Resistivity measurements were carried out at 1023, 1048, 1073, 1098 and 1123 K. The gas switches were performed from $P(O_2) = 10^{-5}$ atm to $P(O_2) = 1$ atm in order to be sure to achieve the desired $P(O_2)$ within the measurements cell instantaneously.

## Results and Discussion

*Theoretical aspects*

The evaluation of the oxygen diffusion was realised by employing the correlation between the oxygen concentration and the electric resistivity of the sample which is a well established procedure to determine the $D$ values[23-26].

In the model applied we consider as a first approximation that the principal diffusion path is the one along the perpendicular axis to the sample surface, *i.e.* the *c*-axis for the film grown on STO (100) and the *a*(*b*)-axis for the one on STO (110). This is a quite reasonable approximation also because the film surface is extremely smooth (roughness of about 0.4 nm) thus suggesting that the lateral diffusion should be largely hindered. So, the diffusion problem can be approximated to a one-dimensional geometry (perpendicular to the sample surface). In the simplest approximation the main diffusion equation is:

$$\left(\frac{\partial c}{\partial t}\right) = D\left(\frac{\partial^2 c}{\partial z^2}\right) \tag{1}$$

which can be solved to give[27,28]:

$$c_z(t,z) = \frac{4c_0}{\pi} \sum_{n=0}^{\infty} \frac{1}{2n+1} e^{-(((2n+1)/2d_f)\pi)^2 Dt} \times \sin\left(\frac{(2n+1)\pi}{2d_f} z\right) \tag{2}$$

where $c_0$ is the initial concentration and $d_f$ is the thin film thickness which for the present case is 50 nm. By integrating from $z = 0$ and $z = d_f$ and for relatively long times[28] only the first part of this equation can be used introducing an error lower than 1%:

$$c_z(t) = \frac{8c_0}{\pi^2} e^{\left[-\pi^2 tD/(4d_f^2)\right]} \tag{3}$$

that can be written as:

$$c_z(t) = \frac{8c_0}{\pi^2} e^{-(t/\tau_c)} \tag{4}$$

where

$$\tau_c = \frac{d_f^2}{\pi^2 D} \tag{5}$$

is the diffusion relaxation time. By assuming proportionality between changes of c and $\rho$[25,29] equation (4) gives the following dependence with resistivity:

$$\frac{\rho(t) - \rho_\infty}{\rho_0 - \rho_\infty} \propto e^{-(t/\tau_c)} \tag{6}$$

where $\rho_0$ is the initial resistivity and $\rho_\infty$ is the saturation resistivity for $t \to \infty$.

Of course, one important parameter in affecting the results of diffusion coefficient determination is the thin film thickness which is directly involved in the pertaining equations. Also the spread of $D$ values usually encountered in the literature for thin films may have this source. In the present case the X-ray reflectivity assures the determination of thin film thickness with a very high precision[30].

Finally, since the oxygen diffusion is expected to be an activated process it is possible to determine the activation energy for that by following the temperature dependence of the diffusion relaxation time:

$$\tau_c(T)^{-1} = \tau_0^{-1} e^{-(E_A/k_B T)} \tag{7}$$

where $E_A$ is the activation energy and $k_B$ the Boltzmann's constant.

*Resistivity Measurements*

As detailed in the experimental section the samples studied are good quality, epitaxial, thin films of $La_{0.85}Na_{0.15}MnO_3$ (LNMO) on STO (100) and (110) with a thickness of 50 nm and a roughness lower than 0.5 nm. Complete details about the physical properties of the deposited films will be presented elsewhere[21]. What is important is that our materials are ferromagnetic already at room temperature since the Curie transition temperature for both films are about 320 K.

Resistivity measurements were collected, on the two films prepared, at 1048, 1073, 1098 and 1123 K. First, the samples were equilibrated in $P(O_2) = 10^{-5}$ atm taking care to attain the gas-solid equilibrium, *i.e.* by following the resistivity variation until a constant value was not reached. Than the gas composition in the cell measurement was changed to $P(O_2) = 1$ atm and the resistivity variation was followed.

Figure 1 reports, as an example, the result of that kind of measurements *vs.* time for the LNMO film grown on STO (100) at 1023 K (empty circles) and for the LNMO thin film grown on STO (110) at 1123 K (empty diamonds). The resistivity data have been normalized taking into account equation (6). As can be appreciated by the plot the time scale for oxygen in-diffusion is very fast and an asymptotical value is reached quite soon. For the film grown on STO (110) at 1123 K a plateau in the resistivity can be now achieved for time shorter than 60 s.

The solid lines in the Figure are the theoretical curves based on equation (6); as can be appreciated they fitted very well the experimental data yielding the relaxation times for the oxygen diffusion. To achieve a good fit of the experimental curves two exponentials were used. So, two different relaxation times and consequently $D$ values were obtained; they are labeled as $\tau'$ and $\tau''$ and $D'$ and $D''$, respectively. In particular, the first region of the relaxation curve (usually around 20 s) seems directly involved in a faster diffusion process which can be accounted for diffusion along twin boundaries. Often this contribution is neglected or not considered since it is a fast process compared to volume diffusion. In the present case the two processes are nearly comparable (even though the twin boundary diffusion is nearly half order of magnitude faster) so a detailed description of the diffusion process has to take it into account.

Figure 2 presents the results of the fitting procedure, that is the $D$ values obtained for the two series of films, while the numerical values are reported in Table 1. The $D''$ diffusion coefficients increase by increasing temperature. On the opposite, The $D'$ values do not show an easy behavior with temperature even though for both samples a common trend can be found. This may came from the complex temperature dependence of grain boundary diffusion[31]. As can be appreciated the $D$ values obtained for the LNMO film grown on STO (100) are greater than the ones obtained for the film on STO (110) for the same temperatures. This is true particularly for the grain boundary diffusion even though also for the bulk one the diffusion values for the film grown on STO (100) are greater of more than 20%-30% with respect to the values for the other film.

Our data, as told before, are the first obtained on oriented thin films of optimally doped manganites. So, a direct comparison with other literature data cannot be performed. Anyway, it is possible to look at the tracer diffusion coefficients determined in the past for undoped and doped manganites[23,31-33]. Unfortunately the scattering of the published $D$ values is big, in some cases greater than six orders of magnitude. It can be seen that our volume diffusion data ($D''$) are in agreement with some results obtained on self-doped $LaMnO_3$[31] and with some Sr-doped samples[33] even though the coefficients determined on polycrystalline samples tend to be from one to two orders of magnitude bigger with respect to the ones determined on epitaxial films.

Figure 3 reports the Arrhenius plots for the data obtained in this study (only the $D''$). Black circles are relative to the LNMO film on STO (100) while the squares refer to the data for the LNMO film on STO (110). In this graph the relaxation rates obtained from the diffusion process and extrapolated by the fitting procedure are plotted towards the reciprocal of temperature. It is clear that the oxygenation process is an activated one. From the plot it is possible to calculate the activation energies. For the film on STO (100) it has a value of about 0.35 eV while for the film on STO (110) it is nearly doubled, 0.66 eV.

## Concluding Remarks

In this paper we have presented the first study devoted to the determination of chemical diffusion coefficient of oxygen in-diffusion in magnetoresistive manganite thin films. Moreover, possible anisotropies in the oxygen diffusion have been taken into account by study two different orientations of thin films, *i.e.* with the long crystallographic axes (*c*) perpendicular or parallel to the substrate surface. The method employed is the isothermal conductivity measurements.

All the time-dependent resistivity data collected have been well analyzed within a diffusion controlled model with the need of two exponential terms, so taking into account also a possible contribution from twin boundaries of thin film grains. The *D* values obtained are lower than the ones published for polycrystalline materials even though a direct comparison is not possible since no works have been reported for a bulk with the same composition of our thin films. We are going to expand our study in the future in order to also include the study of bulk materials.

Interestingly, the oxygen in-diffusion data here determined are in perfect agreement with some data reported for HTSC thin films[26], that is of the order of $10^{-14}$ cm$^2$s$^{-1}$. This is anyway something which could be expected since for both systems the diffusion mechanism is expected to occur via oxygen vacancies since no interstitials oxygens are plausible to occur in perovskite manganites[17]. This is particularly true for optimally doped manganites where a small oxygen under-stoichiometry is also present[7,15].

From our study we could also observe an anisotropy in the oxygen diffusion: the volume diffusion coefficient for the (100) orientation is greater than that for the (110) of few tents percent. Even though it is not expected an ordered oxygen vacancy arrangement for the manganites, as happens for HTSCs, it may be noted that the tendency to form anion vacancies should be partially connected to the tendency to reduce the repulsions of the couple $V_o^{\bullet\bullet}$-Mn$^{3+/4+}$ or -La$^{3+}$/Na$^+$.

Looking at the packing and chemical nature of the different crystallographic faces displayed by the LNMO films grown on STO (100) and STO (110), reported in Figure 4a and 4b, respectively, it is possible to note that, based on a simple electrostatic picture, the first orientation should give rise to a more easier accommodation of the anionic vacancies formed and also to an easier anion movement due to a less dense charge field. At the present stage this figure is an hypothesis on which also modelling treatments will be considered in order to shed light on the difference encountered for the diffusion along the two crystallographic directions.

Finally a comment on the activation energies. Again the energy required for the diffusion in the film grown on STO (100) is lower with respect to the other orientation. This result correlates with the easier migration and formation of oxygen vacancies for the (100) orientation. Interestingly, if we compare our data with the ones determined on HTSC thin films it can be observed that our values are lower. The $E_A$ for oxygen in-diffusion in HTSC are around 1-1.3 eV[25,26]. This discrepancy can be nicely accounted for the basic difference in the nature of the oxygen diffusion mechanism in HTSC (where an *ordered* process takes place with also a well defined sequence of doping[25,34-36]) with respect to manganites where, at least on the basis of the actual experimental available data, no evidences of vacancies (oxygen) ordering processes hold.

## Acknowledgments

Financial support from the Italian Ministry of Scientific Research (MIUR) by PRIN Projects (2002) is gratefully acknowledged.

Dr. Ivano Alessandri and Dr. Valentina Cervetto are kindly acknowledged for thin films preparation.

# Figure captions

**Fig. 1.** Resistivity variation with time for the LNMO film grown on STO (100) at 1023 K (open circles) and at 1123 K (open diamonds).

**Fig. 2.** Diffusion coefficients for the films studied *vs*. *T*. Empty symbols refer to the LNMO film on STO (100), the squares are the *D'* values while the diamond the *D''*. Full symbols represent the *D'* (circles) and *D''* (up-triangle) values for the LNMO film on STO (110).

**Fig. 3.** Arrhenius plot for the LNMO film on STO (100) (circles) and for the LNMO film on STO (110) (squares).

**Fig. 4.** Sketch of manganite cell for the growth on STO (100), 4A, and on STO (110), 4B.

# Table caption

**Table 1.** Temperature and diffusion coefficients for the films studied (see text for details).

# Table 1

| Sample | *T* (K) | *D'* (cm$^2$s$^{-1}$) | *D''* (cm$^2$s$^{-1}$) |
|---|---|---|---|
| LNMO on STO (100) | 1023 | 8.47e-14 | 2.67e-14 |
|  | 1073 | 4.83e-13 | 1.143e-13 |
|  | 1099 | 4.647e-13 | 1.271e-13 |
|  | 1123 | 2.916e-13 | 1.325e-13 |
| LNMO on STO (110) | 1048 | 2.476e-13 | 6.934e-14 |
|  | 1073 | 2.694e-13 | 8.082e-14 |
|  | 1098 | 2.962e-13 | 1.052e-13 |
|  | 1123 | 1.757e-13 | 1.098e-13 |

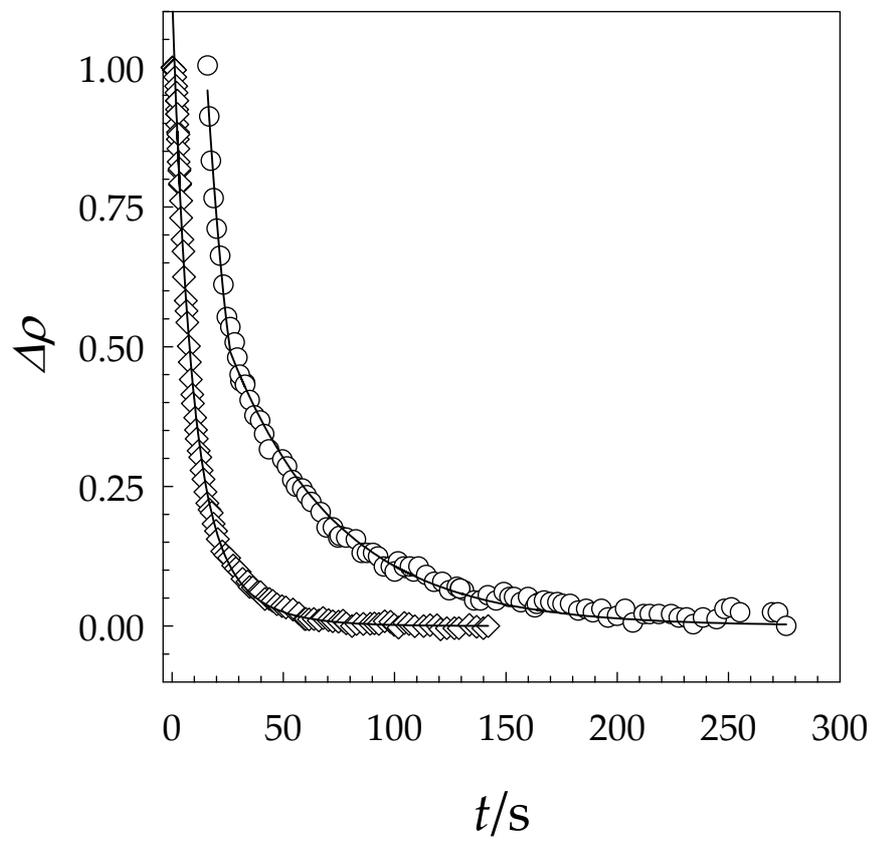

Figure 1

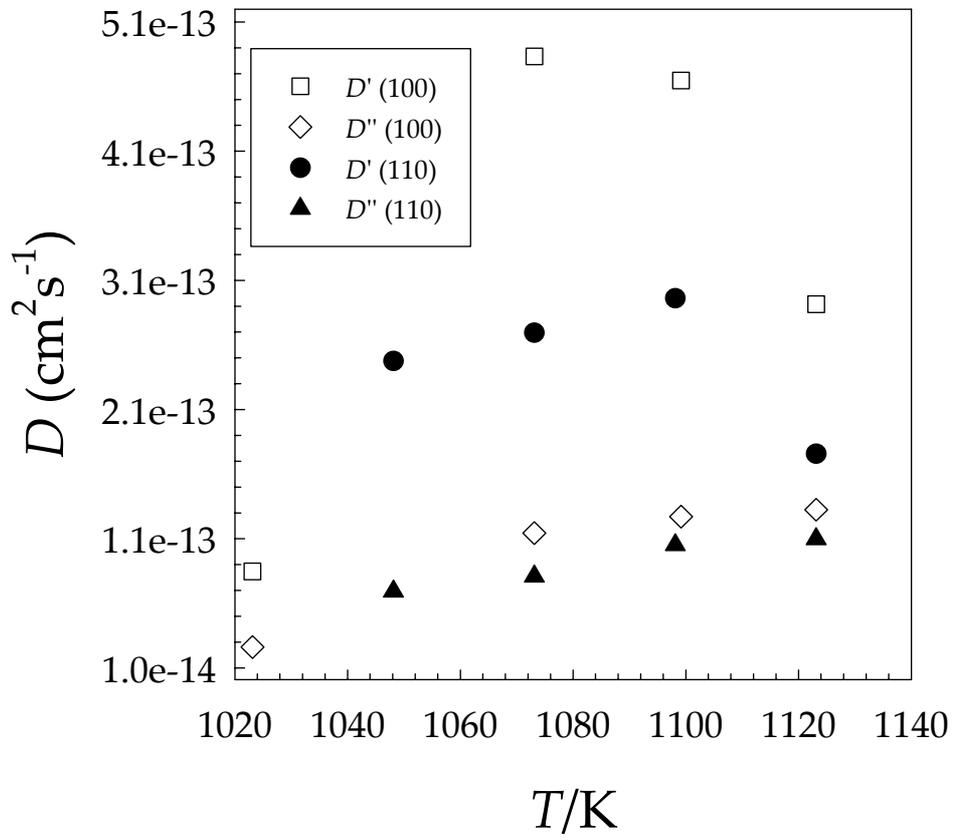

Figure 2

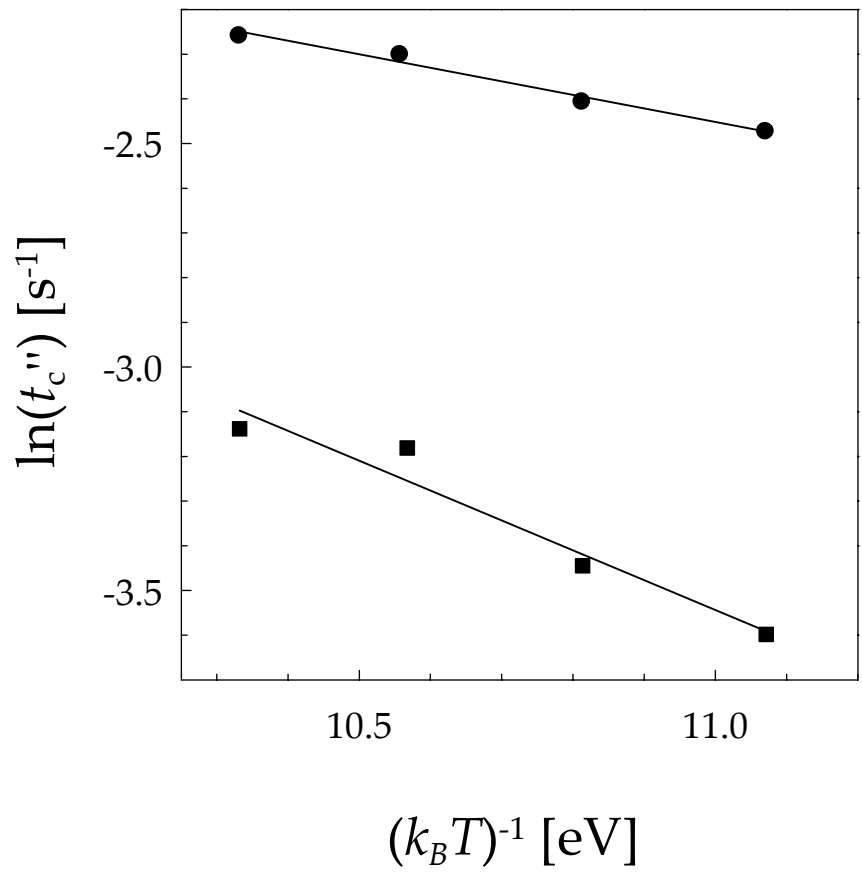

Figure 3

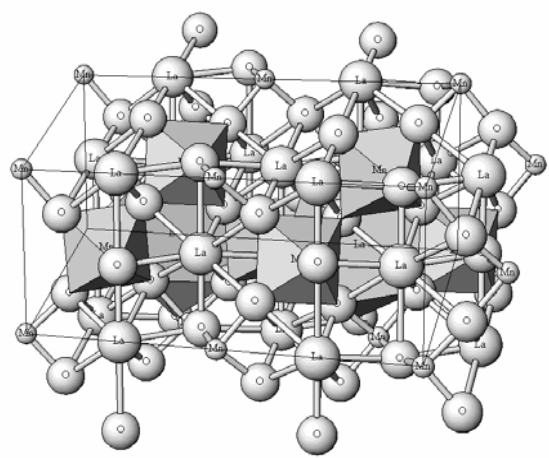

A

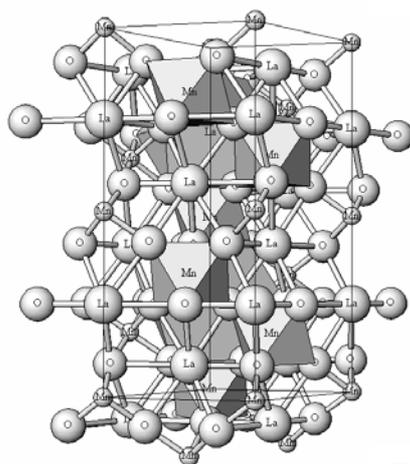
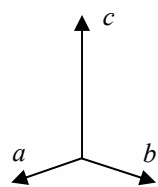

B

Figure 4